\documentclass[]{spie} 

 \pdfoutput=1
\usepackage{amsmath,amsfonts,amssymb}
\usepackage{graphicx}
\usepackage[colorlinks=true, allcolors=blue]{hyperref}

\title{An instrumental puzzle: the modular integration of AOLI}

\author[a,b]{Roberto L. L\'opez}
\author[a,b]{Sergio Velasco}
\author[c]{Carlos Colodro-Conde}
\author[d]{Juan J. F. Valdivia}
\author[a,b]{Marta Puga}
\author[a,b]{Alejandro Oscoz}
\author[a,b,e]{Rafael Rebolo}
\author[f]{Craig Mackay}
\author[c]{Antonio P\'erez-Garrido}
\author[a,b]{Luis Fernando Rodr\'iguez-Ramos}
\author[a,b]{Jos\'e M. Rodr\'iguez-Ramos}
\author[f]{David King}
\author[g]{Lucas Labadie}
\author[g]{Balajai Muthusubramanian}
\author[a,b]{Gustavo Rodr\'iguez-Coira}
\affil[a]{Instituto de Astrof\'isica de Canarias, c/ V\'ia L\'actea s/n, La Laguna, Tenerife E-38205, Spain.}
\affil[b]{Departamento de Astrof\'isica, Universidad de La Laguna, La Laguna, Spain.}
\affil[c]{Universidad Polit\'ecnica de Cartagena, Campus Muralla del Mar, Cartagena, Murcia E-30202, Spain. }
\affil[d]{Departamento de Ingenieria Industrial, Universidad de La Laguna, La Laguna, Spain.}
\affil[e]{Consejo Superior de Investigaciones Cient\'ificas, Madrid, Spain.}
\affil[f]{Institute of Astronomy, University of Cambridge, Madingley Road, Cambridge CB3 0HA, UK.}
\affil[g]{I. Physikalsiches Institut, Universit\"at zu K\"oln, Z\"ulpicher Strasse 77, 50937 K\"oln, Germany}

\authorinfo{Further author information: \\ R.L.L.: E-mail: \href{mailto:rll@iac.es}{rll@iac.es} \\S.V.: E-mail: \href{mailto:svelasco@iac.es}{svelasco@iac.es}
}

\begin{document} 
\maketitle

\begin{abstract}
The Adaptive Optics Lucky Imager, AOLI, is an instrument developed to deliver the highest spatial resolution ever obtained in the visible, ~20 mas, from ground-based telescopes. In AOLI a new philosophy of instrumental prototyping has been applied, based on the modularization of the subsystems. This modular concept offers maximum flexibility regarding the instrument, telescope or the addition of future developments.  
\end{abstract}


\section{INTRODUCTION}
\label{sec:intro} 

The Lucky Imaging (LI) technique offers to ground-based telescopes an excellent and cheap method of reaching diffraction limited spatial resolution in the visible. However, this technique suffers two important limitations: it can be applied only at telescopes with sizes below 2.5m, achieving a resolution similar to that of the HST, and most of the images are discarded, meaning that only relatively bright targets can be observed. AOLI is a state-of-the-art instrument conceived to beat these limitations by combining the two most successful techniques to obtain extremely high resolution, LI and Adaptive Optics (AO). This instrument is hence planned as a double system that includes an adaptive optics closed loop corrective system before the science part of the instrument, this last using LI. The addition of low order AO with a new tomographic image pupil wavefront sensor (TPI-WFS) to the system before the LI camera enhances the reachable resolution as it removes the highest scale turbulence maximizing the LI process at larger telescopes. 

Aiming at this challenging goal we have built AOLI putting together the expertise of several institutions -IAC, IoA, UPCT, UC and ULL-, each group specialized in a different subject, corresponding to a part of the puzzle. To face the defiance that AOLI represents we have implemented a new philosophy of instrumental prototyping by modularizing all its components: simulator/calibrator, deformable mirror (DM), science and WFS modules. This modular concept, which is really suitable for AOLI, offers huge flexibility for changes, such as the hosting telescope or the addition of future developments and improvements. AOLI has now been restructured not only to make the AIV phase reliable but also to be able to integrate this system regarding different parameters (f-number, scale, WFS-type,) or to adapt it to different telescopes. AOLI was initially designed for the 4.2m William Herschel Telescope (WHT, Observatorio del Roque de los Muchachos, La Palma island, Spain); however the excellent results obtained and the versatility obtained through this modular concept will allow us to adapt it to other telescopes, even including the 10.4m GTC (ORM, La Palma island).

Here we will give the answer to major questions about AOLI’s modularity: the why and how of doing it (really well). With such purpose we present a description of the whole instrument, reveal key details of the integration of all its modules and of the verification phase, a crucial step in the development, as well as the preliminary results after its first observing run on the WHT. This work is always done with a double aim in mind: to provide extremely high spatial resolution and to offer a cheap and versatile instrument to be installed on different telescopes.

\section{DESCRIPTION}
\label{sec:aoli} 

AOLI represents the convergence of different techniques explored and extensively developed by teams from different institutions. The Instituto de Astrof\'isica de Canarias (IAC, Spain) leads the project and provides not only the management but also its expertise and capabilities on AIV of state-of-the-art instruments, optical instrumentation and lucky imaging (L\'opez et al.\cite[2014]{2014SPIE.9151E..3ML}, Oscoz et al.\cite[2008]{2008SPIE.7014E.137O}). The Institute of Astronomy (IoA, UK) collaborates with the optics and detectors (Mackay et al.\cite[2014]{2014SPIE.9147E..1TM},\cite[2012]{2012SPIE.8446E..21M}). Universidad Polit\'ecnica de Cartagena (UPCT, Spain) has developed the control and processed images software (Oscoz et al.\cite[2008]{2008SPIE.7014E.137O}), while University of Cologne (Germany) (Labadie et al.\cite[2011]{2011A&A...526A.144L},\cite[2010]{2010SPIE.7735E..32L}) is in charge of the science software. Finally, University of La Laguna (ULL, Spain) contributes with a novel tomography image pupil wavefront sensor (TPI-WFS) ({Rodr{\'{\i}}guez} et al.\cite[2010]{2010aoel.confE5011R}).

  \begin{figure} [ht!]
  \begin{center}
  \includegraphics[height=9cm]{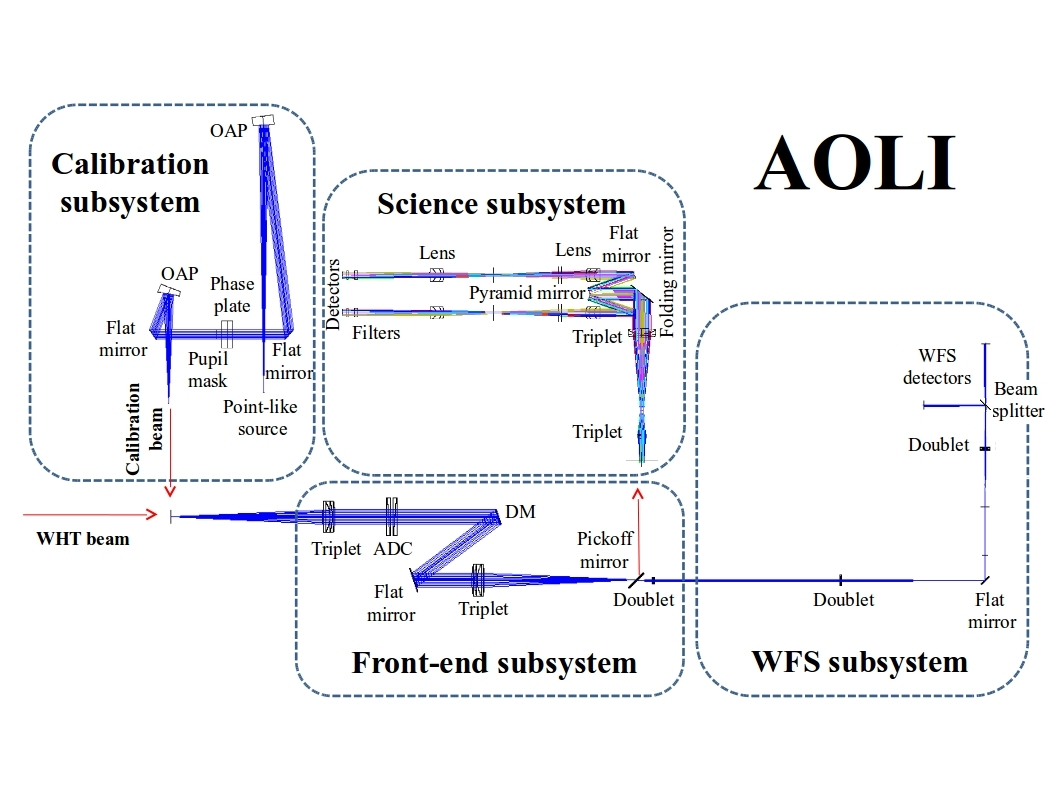}
   \end{center}
  \caption[example] { \label{fig:aoli} AOLI's optical layout. The AO subsystem before the science camera rises the amount of images that can be used for lucky imaging, allowing the observation of much fainter objects.}
  \end{figure} 

The system is composed by four different modules: 
\begin{itemize}
\item Telescope and turbulence simulator and calibrator (SimCal): it delivers a calibrated point-like source resembling the telescope f-ratio and exit pupil and includes turbulence simulation of ground atmospheric layers.
\item Science subsystem based on four synchronized 1kx1k EMCCDs with three different plate scales and field of view (36x36 arcsec to 120x120 arcsec). 
\item Tomographic Pupil Image WFS (TPI-WFS), implemented for the first time on AOLI. It retrieves the pupil by measuring the intensity of defocused pupil images taken at two planes, allowing the use of much fainter reference stars than other WFS systems. The images are sent to two 512x512 pixels EMCCDs.
\item Front-end + AO subsystem: 241 actuators ALPAO deformable mirror (DM) and conditioning optics.
\end{itemize}
 
The TPI-WFS and the AO modules constitute the 4CAOS system (Canarias-Cambridge-Cartagena-Cologne AO System) which can be easily adapted to other telescopes and instruments.

\section{FIRST OPTICAL AND WFS DESIGNS}

This instrument has seen several configurations in an important evolution since the very first design to the successful day of the commissioning, always trying to make it more compact and versatile. During the AIV process we found that a high accuracy in the optical alignment and the repeatability of the mechanical positioners were the clue to succeed. For this purpose, AOLI has been built in blocks over breadboards, which have been integrated and verified independently, each one completely removable and exchangeable. This “puzzle thinking” gives AOLI a high flexibility while keeping its strength allowing to improve and optimize each subsystem independently to mount the final puzzle.

The initial configuration of the instrument was designed for a nonlinear curvature wavefront sensor (nlCWFS) based on the use of two EMCCD 512x512 detectors (Crass et al.\cite[2012]{2012SPIE.8447E..0TC}). Simultaneous images of two pupils were projected on each of these detectors, each of them corresponding to a different band blurred image of the telescope pupil at intra and extra positions from the ideal image plane of a pupil conjugate.

  \begin{figure} [ht]
  \begin{center}
  \includegraphics[height=8cm]{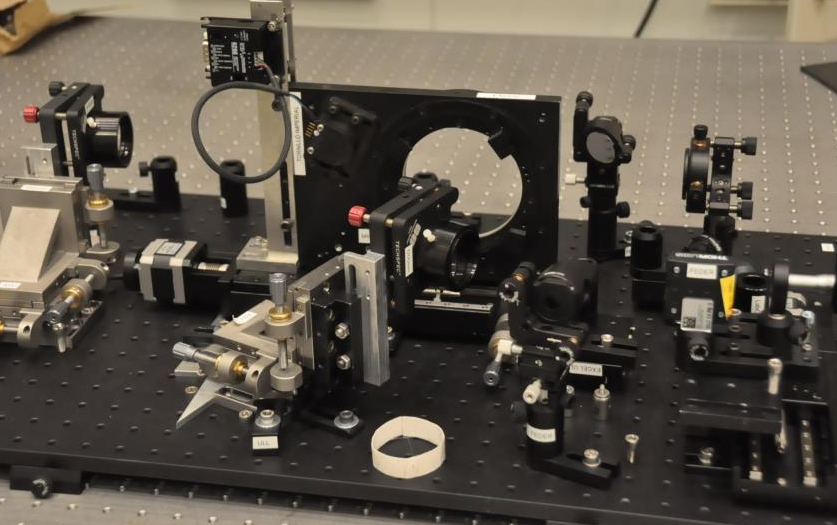}
   \end{center}
  \caption[example] 
  { \label{fig:img1} 
SimCal system achromatic module. Fiber optic (object) placed in top right edge. Bottom right edge showing telescope focus.}
  \end{figure}

A WHT telescope and turbulence simulator (SimCal) (Puga et al.\cite[2014]{2014SPIE.9147E..7VP}) has been built to enable the development and calibration of the instrument at the laboratory. To get rid of the chromatic effects affecting the pupil propagation a preliminary design based on off-axis parabolic (OAP) mirror was implemented.

SimCal consists of a point source generated by an optical fiber of 9μm in diameter illuminated by a halogen lamp. The object is collimated by an off axis parabola and in its conjugate focus we assemble a scaled replica of the WHT's pupil. A rotative pseudo-random phase screen has been placed before the pupil mask in the optical path to simulate low atmospheric layers turbulence. A second OAP refocuses the object with the WHT f-ratio and the size of the equivalent Airy disk. A telecentric configuration has been chosen in order to avoid aberrations and critical effects of pupil alignment on the adaptive optics system, see fig. \ref{fig:img1}.

The focal object generated was verified by a Shack-Hartmann WFS. After the pupil, a set of neutral filters allows to modify the brightness of the object, starting from a calibrated value of apparent magnitude \textit{I} = 6 as obtained with a fixed filter OD = 2.5.
The instrument can be fed at the lab by the simulated beam at a plane equivalent to a telescope post-focal plane, just inserting a motorized positioned flat folding mirror in the telescope-instrument optical path. The first optical component in the instrument optical path is a collimating lens that conjugates the pupil at the DM plane. In order to compensate the atmospheric dispersion, minimizing chromatic effects, a dual Amici prism based ADC is placed within the collimated beam before the DM. The beam emerging from the DM is focused by an achromatic lens on a 45º pick-off mirror. This plate splits the on-axis field light by transmitting around 70\% of it to the WFS and deviating the rest to the science detectors. The off-axis field is almost totally reflected to the science system by a silver coating. The transmitted beam is scaled and extended to two EMCCDs that act as pupil evaluators constituting the WFS.

  \begin{figure} [ht]
  \begin{center}
  \includegraphics[height=8cm]{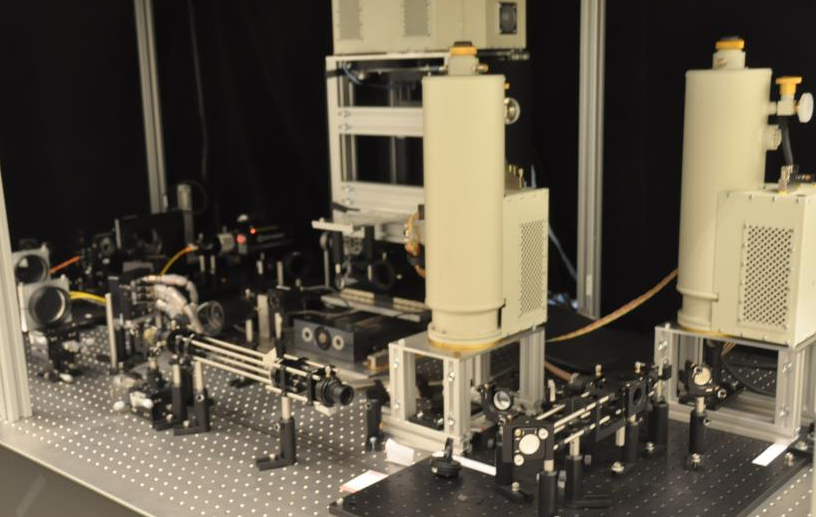}
   \end{center}
  \caption[example] { \label{fig:img2} Original setup for the second commissioning: WFS (front), science subsystem (center), and collimator and camera lenses together with the DM (left). 
}
  \end{figure} 

In the original format the beam was divided by a dichroic mirror that splits the light into two bands, B+V and R+I, which were further divided and sampled at four different defocused intra and extra pupil planes by means of two additional dichroics.

This setup was the base of the the nlCWFS configuration (McCay et al.\cite[2014]{2014SPIE.9147E..1TM}, Crass et al.\cite[2015]{2015AAS...22541306C}) with which took place the first commisioning at WHT on September 2013. The extremely complicated nlCWFS and the weather conditions did not allowed the use of AO and only Lucky Imaging data was obtained during that run (Velasco et al.\cite[2016]{2016MNRAS.tmp..855V}).

\section{RECONFIGURATION: THE PUZZLE CONCEPT}

In spite of the problems found during this first partial commissioning, very valuable science images were taken and the potential of the instrument was demonstrated. However, two problems affecting the instrument were evident:
\begin{itemize}
\item We realized how difficult and tedious was to assemble the optical system on the GHRIL optical bench in the WHT Nasmyth focus.
This was confirmed in the subsequent testing and adjustment period at the IAC lab: the optical alignment was very sensitive to temperature changes and the optomechanical components showed a drift behaviour from the nominal positions. The lever arms of the optical system were too large, leading to misalignments, and the science camera system  was heavy and bulky, making it difficult to align. The DM position was not static due to the stiffness of the wiring. 
\item It was almost impossible to solve the phase mapping algorithm with this nlCWFS.
\end{itemize}

The consequences of these findings were crucial for the final success of AOLI: the modular concept was adopted and we finally decided to use an adaptation of the tomographic phase reconstruction system for 3D images developed by the ULL ({Rodr{\'{\i}}guez-Ramos} et al.\cite[2012]{2012SPIE.8384E..10R}, \cite[2011]{2011SPIE.8043E..25R}). 

The main characteristic of the TPI-WFS is that, to reconstruct the phase map by obtaining a Zernike polynomials decomposition up to a high order, two defocused images of the pupil are obtained. One of the advantages of the new system is that now we have to deal with only two images, one on each detector, and in white light Which provides much better use of the photons per unit area as their intensity is divided by two instead of by four. To validate and compare the retrieved wavefront we counted 
with a commercial SH-WFS at a pupil conjugate in the simulator subsystem. 

  \begin{figure} [ht]
  \begin{center}
  \includegraphics[height=8cm]{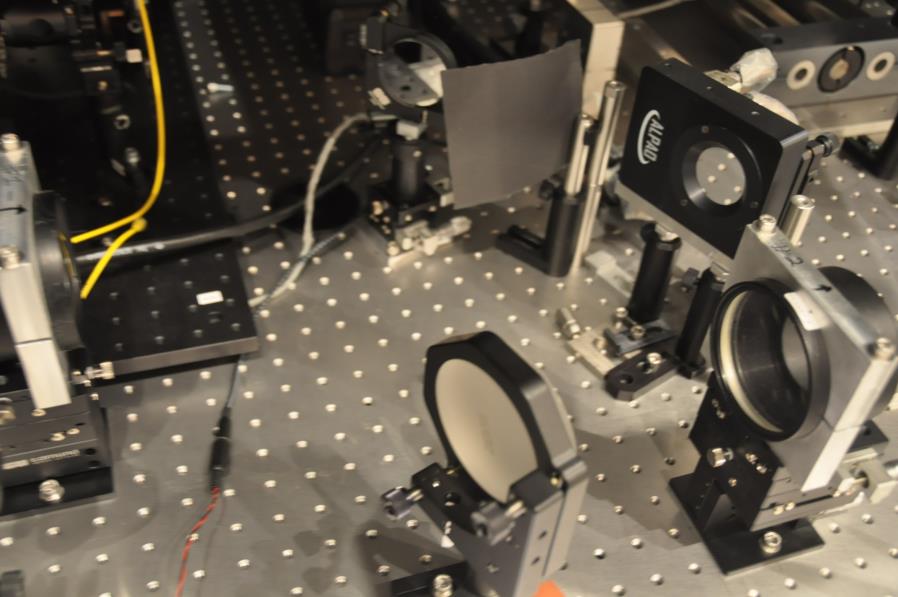}
   \end{center}
  \caption[example] { \label{fig:img3} Original DM system without ADC. The angle between incoming and out-coming DM beams is 36 degrees.}
  \end{figure} 
  
In addtion, and in order to minimize the number of components to build up and align in every instrument assembly, we considered a modular system configuration. On the one hand we separated the simulator and calibration system and mounted it on a small optical breadboard placed on rails, so that we could move it as a block and adjust its position finely on the rails.

The stability obtained in the SimCal with this procedure encouraged us to follow a similar scheme in other subsystems. The SimCal insertion mirror, the collimator lens, ADC, the DM, the folding mirror, and the camera lens were mounted on another optical breadboard in a much more compact configuration. The axis of this subsystem was aligned with the telescope axis, defined by a line of holes in the optical bench.

  \begin{figure} [ht]
  \begin{center}
  \includegraphics[height=8cm]{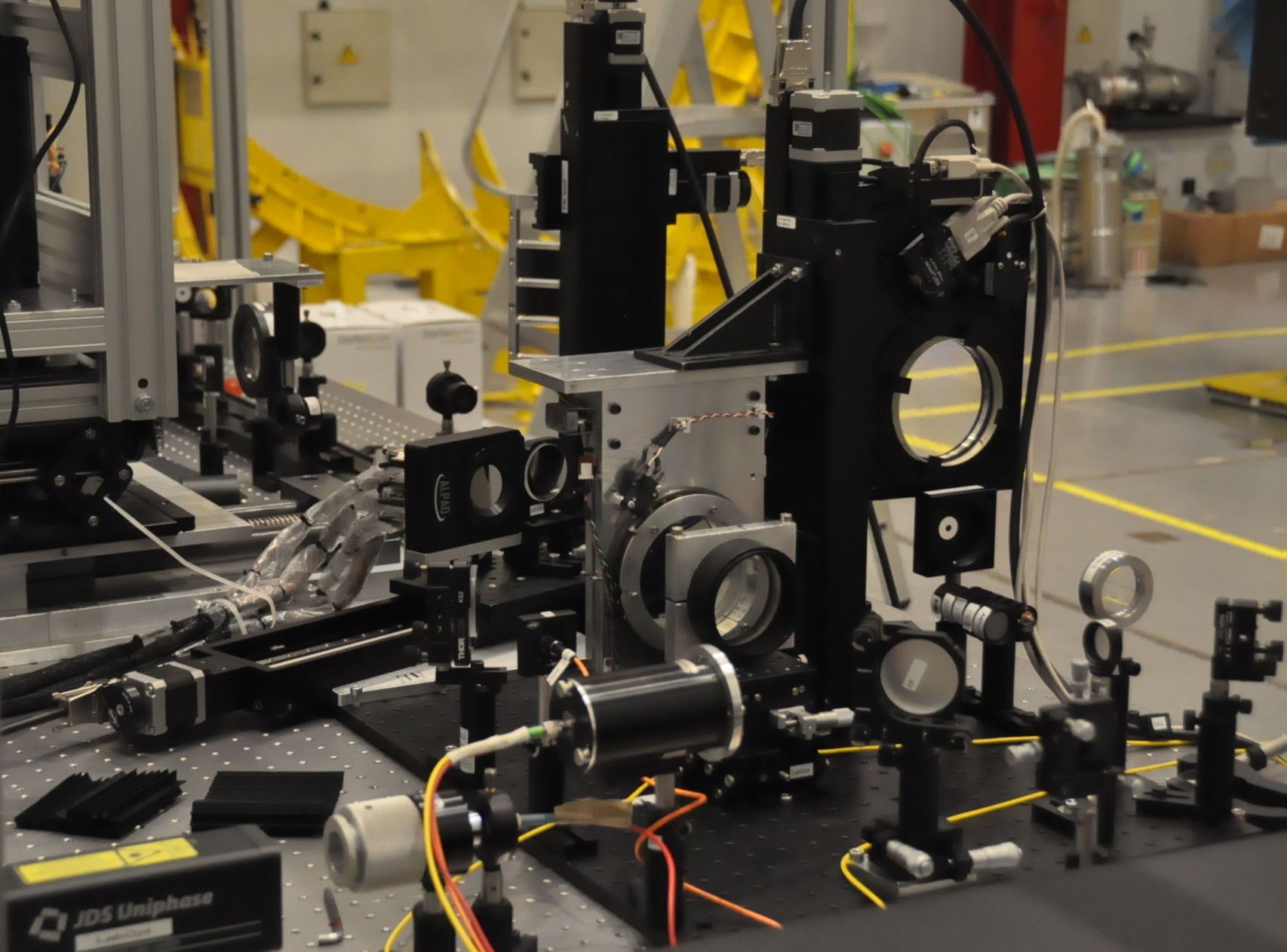}
   \end{center}
  \caption[example] { \label{fig:img4} New DM module with ADC and the last most compact calibration system incorporated to the optical bredboard}
  \end{figure} 
 
At this point, a new modification had to be taken into account for the second commissioning: part of the new instrument for the WHT, WEAVE, will be installed at GHRIL and hence AOLI had to be moved to the other Nasmyth focus, GRACE. This focus is optimized for the NIR band, more convenient for our EMCCD detectors given their quantum efficiency curve. Thanks to the module based configuration of the instrument this relocation did not imply major changes in the instrument configuration. This was indeed the first test of AOLI's versatility.

The changes in the design had an effect on the other subsystems too. 

The WFS subsystem was simplified, installing lenses of shorter focal distances,
and mounting it on a linear board to be aligned independently, leading to a very stable setting. In addition, the detectors with the cryostats were mounted on two bases that we could easily handle, align and fix to the optical table. The science camera constituted a subsystem itself, including an optical scale changer and a FoV re-imaging system onto the four sensors in a mosaic configuration. The Pick-Off Mirror (Poff or POM) is the interface element linking the DM module, the WFS module and the science module. Its configuration provides XYZ movement to focus (z) and position (x, y) the aperture (hole) to the WFS. This system is a separate module with certain degrees of freedom for positioning, easily calibrated.

\section{FINAL MODIFICATIONS AND VERIFICATION PHASE}

In order to further optimize the instrument, we also modularized its software as we did with the optical subsystems, preventing further reconfigurations:
\begin{itemize}
\item	A module for open loop control of the DM depending on the optimization of the science image quality was developed, besides of the closed loop control of the DM based on reconstruction parameters of WFS.
\item	The WFS reconstruction module, was capable of running on both circular and annular pupil images. 
\item	Image acquisition control module for the WFS, communicated with the WF re-constructor stage.
\item	Control module of the linear stages, motors and actuators. 
\item	Science module capable of reading 4 sensors and reconstructs the science FOV by  firstly applying image preprocessing based on LI system from FastCam (Oscoz et al.\cite[2012]{2012SPIE.8384E..10R})
\item	Supervisor framework that controls and communicates with the whole instrument. 
\item	Post processing modle, in a separate package,for the finest treatment of acquired and stored data.
\end{itemize}

All the tests and software carried out during the last phase of the instrument were based on this modular configuration, allowing us to absorb all changes of configuration coming up.

With the simplification of WFS and DM modules, we turn to prepare a simple science module consisting of a single detector but working in classical photometric mode of EMCCD image to LI. On the other hand, as we only had one other equivalent detector, also EMCCD, smaller and allowing a reading speed of images in a few milliseconds, we designed a new compact system through a lateral prism to generate a double image of pupil on the detector, both around of the nominal focus of pupil image. This system was inserted in a simple manner in the way the original WFS by a lateral prism that pulled out the two displaced/delayed beams to the new TFI-WFS module. The multiple alternative system was also more compact and hardy.

  \begin{figure} [ht]
  \begin{center}
  \includegraphics[height=8cm]{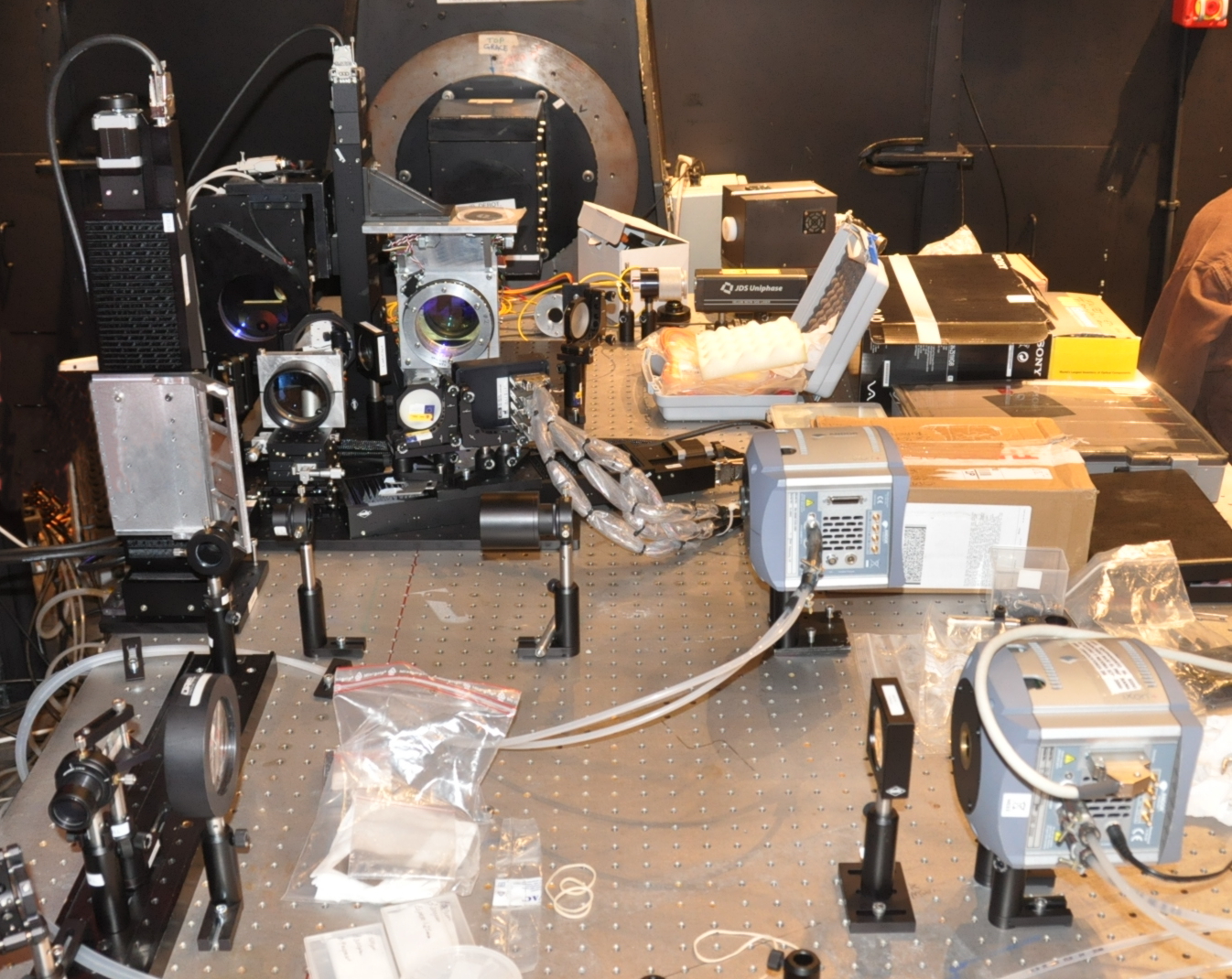}
   \end{center}
  \caption[example] { \label{fig:img5} The AOLI final configuration in Nasmyth GRACE focus of WHT during its integration.}
  \end{figure} 

\begin{figure} [ht]
  	\begin{center}
  		\includegraphics[height=5cm]{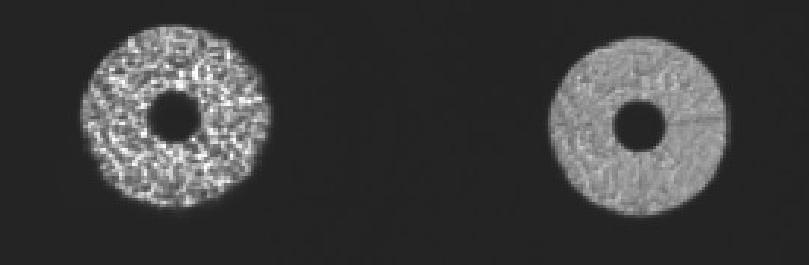}
  	\end{center}
  	\caption[example] { \label{fig:img6} TPI-WFS use two defocused pupil images away pupil ideal plane.}
  \end{figure}

The first commissioning of the full instrument took place at WHT on May 21st, 2016. The weather conditions that night, with a seeing around 2.2 arcsec, well above what is generally taken as a limit for AO systems, and full Moon, were not optimal for high spatial resolution techniques. However, we were able to close the loop with stars of different magnitudes making use of the TPI-WFS and an ALPAO deformable mirror with 244 actuators. In addition, the system was fed with a calibration made in real time from the image of the pupil (see fig. \ref{fig:img6}) from the twinkling star itself.

AOLI is, hence, the first instrument to succeed in closing the loop with stellar sources using the novel TPI-WFS. 

  \begin{figure} [ht]
  	\begin{center}
  		\includegraphics[height=8cm]{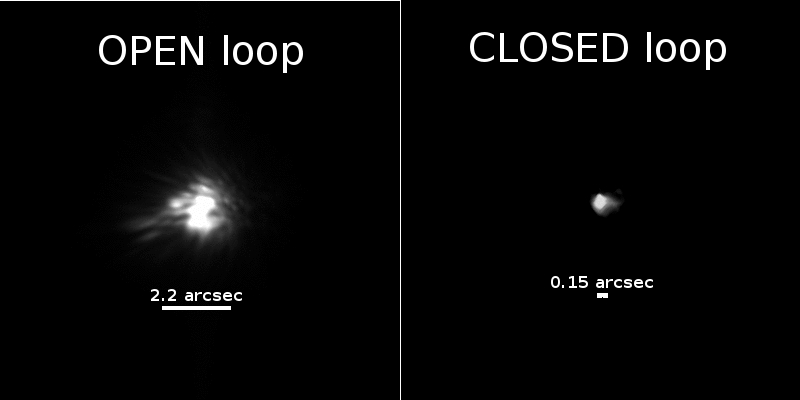}
   	\end{center}
  	\caption[example] { \label{fig:img7} The AO closed loop was obtained in not good seeing conditions for NIR band.}
  \end{figure} 


\section{CONCLUSIONS AND FURTHER WORK}

As a consequence of the modularity of its components, AOLI@WHT is much smaller and more efficient than its first design. This success has lead us to plan to condense it even more to get a portable and easy to integrate system, ALIOLI (Adaptive and Lucky Imaging Optics Lightweight Instrument). It will consist of a DM+WFS module with a lucky imaging science camera attached, and the SimCal will be used as an independent module to verify the alignment and calibration of this or similar instruments. 

Our aim is to install ALIOLI at some of the telescopes of the Observatorios de Canarias. Furthermore this modularized concept permits to install not only a LI camera as science detector but any independent instrument or science detector benefiting from the AO capabilities.  We also have designed an optimized solution to bring AOLI to GTC.

\section{MULTIMEDIA FIGURES}

You can see some images, animations and new data over this instrument in \href{http://www.iac.es/proyecto/AOLI/}{IAC AOLI project homepage} [\url{http://www.iac.es/proyecto/AOLI/}]

\acknowledgments 

This paper is based on observations made with the William Herschel Telescope operated on the island of La Palma by the Isaac Newton Group in the Spanish Observatorio del Roque de los Muchachos of the Instituto de Astrof\'isica de Canarias

\bibliography{main} 
\bibliographystyle{spiebib} 

\end{document}